\documentclass[10pt, conference]{IEEEtran}
\IEEEoverridecommandlockouts
\usepackage{cite}
\usepackage{amsmath,amssymb,amsfonts}
\usepackage{subfigure}
\usepackage{algorithmic}
\usepackage{graphicx}
\usepackage{textcomp}
\usepackage{xcolor}
\usepackage{booktabs} 
\usepackage{soul}
\graphicspath{{figs/}}
\usepackage{braket}
\usepackage{etoolbox}
\usepackage{fancyhdr}
\usepackage{multirow}
\def\BibTeX{{\rm B\kern-.05em{\sc i\kern-.025em b}\kern-.08em
    T\kern-.1667em\lower.7ex\hbox{E}\kern-.125emX}}
\begin{document}

\title{STIQ: \underline{S}afeguarding \underline{T}raining and \underline{I}nferencing of \underline{Q}uantum Neural Networks from Untrusted Cloud}

\author{\IEEEauthorblockN{Satwik Kundu}
\IEEEauthorblockA{Pennsylvania State University\\
University Park, USA \\
sxk6259@psu.edu}
\and
\IEEEauthorblockN{Swaroop Ghosh}
\IEEEauthorblockA{Pennsylvania State University\\
University Park, USA \\
szg212@psu.edu}
}

\maketitle

\begin{abstract}
The high expenses imposed by current quantum cloud providers, coupled with the escalating need for quantum resources, may incentivize the emergence of cheaper cloud-based quantum services from potentially untrusted providers. Deploying or hosting quantum models, such as Quantum Neural Networks (QNNs), on these untrusted platforms introduces a myriad of security concerns, with the most critical one being model theft. This vulnerability stems from the cloud provider's full access to these circuits during training and/or inference. In this work, we introduce STIQ, a novel ensemble-based strategy designed to safeguard QNNs against such cloud-based adversaries. Our method innovatively trains two distinct QNNs concurrently, hosting them on same or different platforms, in a manner that each network yields obfuscated outputs rendering the individual QNNs ineffective for adversaries operating within cloud environments. However, when these outputs are combined locally (using an aggregate function), they reveal the correct result. Through extensive experiments across various QNNs and datasets, our technique has proven to effectively masks the accuracy and losses of the individually hosted models by upto 76\%, albeit at the expense of $\leq 2\times$ increase in the total computational overhead. This trade-off, however, is a small price to pay for the enhanced security and integrity of QNNs in a cloud-based environment prone to untrusted adversaries. We also demonstrated STIQ's practical application by evaluating it on multiple real quantum hardwares, showing that STIQ achieves up to $\approx$ 70\% obfuscation, with combined performance similar to an unobfuscated model.
\end{abstract}


\section{Introduction}
Quantum computing is rapidly progressing, with companies like Atom Computing and IBM recently unveiling the largest quantum processors ever developed, boasting 1,225 and 1,121 qubits, respectively \cite{computing2023quantum, gambetta2023hardware}. The significant interest in quantum computing among academic and research communities stems from its potential to offer substantial computational speedups over classical computers for certain problems. Researchers have already begun leveraging these noisy intermediate-scale quantum (NISQ) machines to demonstrate practical utility in this pre-fault-tolerant era \cite{kim2023evidence}. Within this emergent field, quantum machine learning (QML) has also gained considerable attention, merging the power of quantum computing with classical machine learning algorithms. QML heuristically explores the potential of improving learning algorithms by leveraging the unique capabilities of quantum computers, opening new horizons in computational speed and capability. Several QML models have been explored, including quantum support vector machines (QSVMs) \cite{rebentrost2014quantum}, quantum generative adversarial networks (QGANs) \cite{dallaire2018quantum}, and quantum convolutional neural networks (QCNNs) \cite{cong2019quantum}. However, quantum neural networks (QNNs) \cite{schuld2014quest, abbas2021power} stand out as the most notable development, mirroring the structure and function of classical neural networks within a quantum framework.

\begin{figure}[!t]
        \vspace{0mm}
        \centering 
        \includegraphics[width=\linewidth]{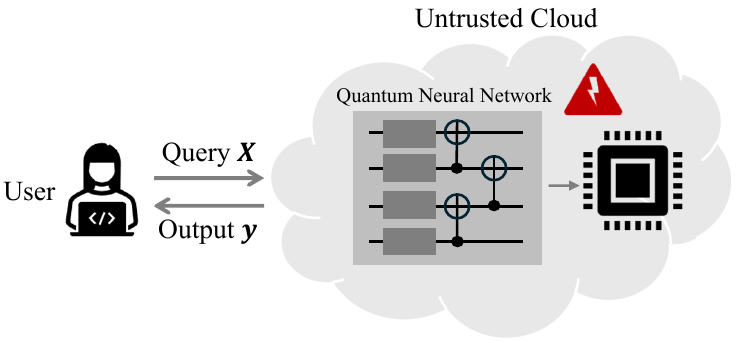}
        \vspace{-4mm}
        \caption{Figure depicting the workflow of QMLaaS, showcasing how users can access QNNs through API queries. Deployment of such QNNs on untrusted clouds exposes them to various potential adversarial attacks.}
        \label{untrusted_cloud}
        \vspace{-4mm}
\end{figure}

At the core of QNNs lies the parameterized quantum circuit (PQC), which consists of several trainable single-qubit and two-qubit gates. The parameters of these gates are updated using classical optimizers to solve the required task. PQCs have been successfully employed in various algorithms to tackle a wide range of problems, including combinatorial optimization \cite{farhi2014quantum}, molecular energy calculation \cite{kandala2017hardware}, classification \cite{farhi2018classification}, finance \cite{nakaji2022approximate} and security \cite{beaudoin2022quantum, kundu2022security}. However, optimizing PQCs, especially in the context of QNNs, presents significant challenges in terms of time, complexity, and quantum resource requirements. This is because, no-cloning theorem, a fundamental principle of quantum mechanics, prevents the duplication of quantum states, making traditional backpropagation techniques used in classical neural networks impractical. To circumvent this limitation, alternative methods like the parameter shift rule are employed which requires executing the quantum circuit multiple times to estimate the gradients needed for optimization. Specifically, each parameter in the circuit needs to be run twice for gradient computation, significantly increasing the quantum resource demand as the number of parameters grows. Thus, as QML expands, we will soon see a rising trend towards hosting uniquely architected QML models or those trained on novel quantum datasets on the cloud (Fig. \ref{untrusted_cloud}), with paid access leading to the rise of Quantum Machine-Learning-as-a-Service (QMLaaS).

\subsection{Motivation}

As access to current quantum hardware remains limited and extremely expensive, it paves the way for the emergence of third-party quantum cloud providers, which could improve access to quantum computers, but potentially at the cost of compromised security. As a result, even though QMLaaS could democratize access to cutting-edge PQC based models, it also exposes QML models to several adversarial attacks 
A particularly pressing concern is the risk of "model theft" attacks. In these scenarios, attackers can steal a model's architecture and behavior because hosting models on untrusted cloud platforms can provide adversaries with white-box access. This means they have complete insight into the specifics of the QNN circuit like the trained parameters, architecture, etc. Consequently, when a QNN is sent to the cloud for training or hosting, the cloud provider might potentially access and replicate the model. This not only jeopardizes the uniqueness of the QNN but also poses a significant risk of intellectual property theft. The provider could potentially exploit this access to offer a similar service, leveraging the stolen model. Given these risks, there is a need for robust security measures when employing untrusted cloud services for QNNs to prevent against exploitation and safeguard intellectual property.

Several studies have addressed protecting quantum circuits from untrusted compilers \cite{saki2021split, suresh2021short} or clouds \cite{upadhyay2022robust}. However, there's a notable gap in research regarding the protection of QNNs from cloud-based adversaries during training and inference stages. Techniques designed for untrusted compilers are inadequate for cloud scenarios, as they only secure the circuit pre-compilation, allowing cloud-based adversaries full access thereafter. Moreover, while quantum homomorphic encryption (QHE) \cite{dulek2016quantum, fisher2014quantum} offers theoretical security, it remains impractical due to its significant overhead. Blind quantum computing (BQC) \cite{broadbent2009universal, barz2012demonstration} requires clients to perform specific operations on a quantum hardware locally, which is often unfeasible. Authors in \cite{patel2023toward} suggested adding random X gates to quantum circuits to obfuscate outputs, but this method has vulnerabilities. Determined adversaries could reverse-engineer the circuit by testing various combinations, and patterns in output distribution might reveal the placement of X gates. Classical techniques for protecting Deep Neural Networks (DNNs) from untrusted cloud environments \cite{chen2018securenets, xu2020secure, juvekar2018gazelle} typically employ matrix transformation or encryption-based strategies. However, these methods are not directly transferrable to QNNs since, the measurement process in QNNs is irreversible, which complicates the application of classical defense strategies that rely on reversible transformations. Furthermore, the quantum counterparts of these encryption techniques tend to introduce significant overhead, making them impractical. This underscores the urgent need to develop more effective and practical protocols to safeguard QNNs from untrusted cloud providers.

\subsection{Contribution}

Here, we introduce STIQ (Safeguarding Training and Inferencing of QNNs), a novel approach designed to safeguard QML models against model theft by untrustworthy cloud providers. Our technique draws inspiration from the principles of ensemble learning but differs fundamentally in the training methodology. Unlike traditional ensemble learning, where each model is independently trained to achieve accurate predictions before their outputs are merged to produce the final result, STIQ trains models in parallel, ensuring that each generates obfuscated results with deliberately reduced individual performance. However, these outputs, when aggregated, reconstruct the accurate final prediction. This method ensures that even though untrusted cloud providers may access the parameters and architecture of the QNNs, the data remains protected, as the outputs of individual QNNs are erroneous and misleading to potential adversaries.   
Our evaluation of STIQ spans various datasets and hardware configurations, showcasing its simplicity and effectiveness in defending QNNs against potential threats posed by cloud-based adversaries. We also tested STIQ on real quantum hardwares (IBM\_Sherbrooke, IBM\_Brisbane and IBM\_Kyoto) and demonstrated the effectiveness of our approach in a more realistic scenario.


\section{Background} \label{background}

\subsection{Basics of Quantum Computing} A quantum bit or qubit is a fundamental building block of quantum computers and is usually driven by microwave or laser pulses for superconducting and trapped-ion qubits, respectively. Unlike a classical bit, a qubit can be in a superposition state, which is a combination of states $\ket{0}$ and $\ket{1}$ at the same time. Mathematically, a qubit state is represented by a two-dimensional column vector $\begin{bmatrix} \alpha \\ \beta \end{bmatrix}$ where $|\alpha|^2$ and $|\beta|^2$ represent probabilities that the qubit is in state `0' and `1', respectively. Quantum gates are operations that change the state of qubits, allowing them to perform computations. Mathematically, they can be represented using unitary matrices 
. There are mainly two types of quantum gates: 1-qubit (like H, X gates) and 2-qubit (like CNOT, CRY gates). Complex 3-qubit gates like Toffoli gates are eventually broken down into 1-qubit and 2-qubit gates during the compilation process.Qubits are measured on a desired basis to determine the final state of a quantum program. Measurements in physical quantum computers are typically restricted to a computational basis, such as the Z-basis in IBM quantum computers. Due to the high error rate of quantum computers, obtaining an accurate output after measuring just once is unlikely. Consequently, quantum circuits are measured multiple times ($n$), and the most probable outcomes from these measurements are then considered as the final output(s). 

\subsection{Quantum Neural Network (QNN)} 
QNN mainly consists of three building blocks: (i) a classical to quantum data encoding (or embedding) circuit, (ii) a parameterized quantum circuit (PQC) whose parameters can be tuned (mostly by an optimizer) to perform the desired task, and (iii) measurement operations. There are a number of different encoding techniques available (basis encoding, amplitude encoding, etc.) but for continuous variables, the most widely used encoding scheme is angle encoding where a variable input classical feature is encoded as a rotation of a qubit along the desired axis \cite{abbas2021power}. As the states produced by a qubit rotation along any axis will repeat in 2$\pi$ intervals, features are generally scaled within 0 to 2$\pi$ (or -$\pi$ to $\pi$) in a data pre-processing step. In this study, we used $RZ$ gates to encode classical features into their quantum states.

\begin{figure}[!t]
        \vspace{0mm}
        \centering 
        \includegraphics[width=\linewidth]{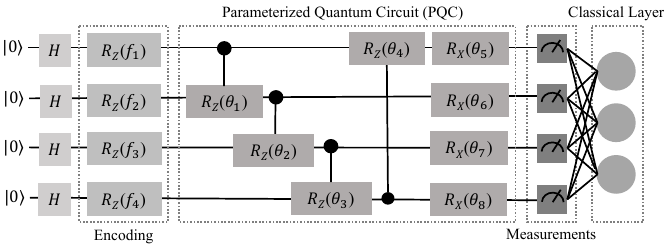}
        \vspace{-4mm}
        \caption{Architecture of a 4-qubit hybrid QNN. Classical features are encoded as angles of quantum rotation gates ($R_Z$). PQC transforms encoded states to explore the search space and entangle features. Measured expectation values are then fed into a classical linear layer for final prediction.}
        \label{qnn_circuit}
        \vspace{-4mm}
\end{figure}

A PQC consists of a sequence of quantum gates whose parameters can be varied to solve a given problem. In QNN, the PQC is the primary and only trainable block to recognize patterns in data. The PQC is composed of entangling operations and parameterized single-qubit rotations. The entanglement operations are a set of multi-qubit operations (may or may not be parameterized) performed between all of the qubits to generate correlated states and the parametric single-qubit operations are used to search the solution space. 
Finally, the measurement operation causes the qubit state to collapse to either `0' or `1'. We used the expectation value of Pauli-Z to determine the average state of the qubits. The measured values are then fed into a classical neuron layer (the number of neurons is equal to the number of classes in the dataset) in our hybrid QNN architecture as shown in Fig. \ref{qnn_circuit}, which performs the final classification task.

\section{Related Works and Limitations}
Model theft poses a substantial security risk to emerging QML platforms particularly with the advent of third-party quantum cloud providers, where even QNN circuits submitted for training and/or inference are vulnerable. To address the generic quantum circuit security risks, strategies have been explored, such as by authors in \cite{upadhyay2022robust}, which suggests dividing circuit execution between trusted and untrusted providers to mitigate adversarial interference risks. However, this approach proves ineffective for QNNs, as access by any single untrusted provider could lead to model theft. Cryptography-based techniques like quantum homomorphic encryption (QHE) offers a theoretical solution by enabling computation on encrypted quantum data, thereby safeguarding the privacy of the computation outcomes \cite{yu2014limitations, dulek2016quantum, fisher2014quantum}. However, the exponential increase in computational overhead, render it impractical in the near term. Likewise, blind quantum computing (BQC) \cite{broadbent2009universal, barz2012demonstration, mantri2013optimal} allows a user to perform quantum computations on a remote quantum computer without revealing the nature of the computation to the server. However, it still requires the client to have the capability to prepare specific quantum bits and perform certain quantum operations which poses a barrier for several users.

There have also been obfuscation based strategies to protect quantum circuits from less trustworthy cloud providers \cite{patel2023toward, upadhyay2022robust, kundu2024evaluating} and compilers \cite{suresh2021short}. Despite these efforts, to the best of our knowledge, there have not yet been any methods developed that effectively protect QNNs against model theft. This is a significant concern because circuit obfuscation techniques, while potentially useful in some cases, cannot be straightforwardly applied to QNNs due to several inherent drawbacks. Specifically, in \cite{patel2023toward} authors proposed randomly adding X gates at the end of the quantum circuit to obfuscating outputs from adversaries. However, an adversary with sufficient time and resources can systematically brute-force the outputs. Furthermore, patterns in the output distribution or correlations between different runs of the circuit could provide hints about the underlying placement of X gates. 


Recent works in the classical domain have addressed secure inference of neural networks on untrusted clouds. For instance, the authors in \cite{chen2018securenets} discuss a method involving the multiplication of neural network data and weight matrices by pseudorandom sparse matrices. This manipulation allows secure execution on the cloud, with subsequent local computations retrieving the correct results. However, this technique cannot be directly applied to QNNs due to the irreversible nature of the measurement process in quantum computing, which complicates the retrieval of correct outputs using classical computations. Additionally, while there have been efforts to implement encryption-based methods \cite{juvekar2018gazelle, xu2020secure, liu2017oblivious}, these approaches are challenging to integrate into quantum circuits. Those that can be adapted incur significant overheads. These dilemmas highlights the challenge in safeguarding QNN circuits against model theft without compromising their operational integrity.

\section{Proposed Technique and Implementation}
Here, we start with the QNN assets and threat model. Subsequently, we provide a brief high-level overview of the proposed technique and a discussion of our training and inferencing methodologies.

\subsection{QNN Assets and Adversary Motivation}
QNN architecture contain myriad of private assets namely, number of layers, entanglement architecture, number of parameters and type of measurement basis, to name a few. These features are decided after significant research, trials and time. Furthermore, the intermediate parameter values during training and final parameter values during inferencing can be considered as assets since they take significant time and cost to obtain. QNNs also embed training data and/or private inferencing data which are additional assets but STIQ is not geared towards protecting them. Adversary will be motivated to steal the QNN and/or its assets to avoid paying for (i) the time and resources needed to design a QNN from scratch, (ii) the training data and (iii) the time and resources needed for training the model. 
\subsection{Threat Model}
Deploying QNNs on cloud services introduces significant security concerns, particularly when these services are managed by entities that might not be entirely trustworthy. This situation presents a specific threat model, where the cloud provider is considered a potential adversary. The main issue is that the trained QNN circuits function as a "white-box" models to the cloud provider, which means that the cloud-provider essentially has full access to the QNN assets e.g., the trained parameters and the architecture. 
When a QNN is sent to the cloud for training or hosting, the provider can simply steal/replicate the model. This not only threatens the QNN's uniqueness but also risks intellectual property theft, as the provider could offer a similar or identical service based on this stolen model. 
The stealth could allow competitors to gain insights into victim's business processes or technological innovations. This scenario highlights the critical need for security measures when working with cloud services for sensitive technologies like QNNs, to protect against potential exploitation and intellectual property infringement.

\begin{figure}[!t]
        \vspace{0mm}
        \centering 
        \includegraphics[width=\linewidth]{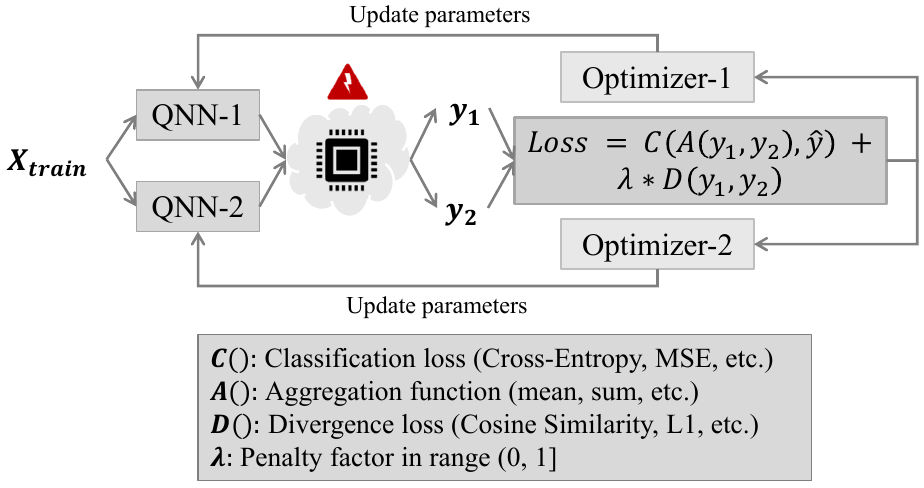}
        \vspace{-4mm}
        \caption{Figure illustrating the training workflow of STIQ. Both QNN-1 and QNN-2 are trained using a unified loss function that optimally balances classification loss $C()$ and divergence loss $D()$.}
        \label{training}
        \vspace{-4mm}
\end{figure}

\subsection{Methodology Overview}
Ensemble learning, recently adapted for the quantum domain, involves training QNNs across various quantum hardware platforms.
Each model makes an independent prediction, and the ensemble model selects the output from the quantum hardware demonstrating the highest confidence. Despite each model's aim for accurate predictions, hosting them on an untrusted cloud exposes them to potential theft. 
Our approach involves training models parallelly on either identical or varied quantum hardware, which is distributed across either the same or different cloud providers. The key idea lies in the models producing highly erroneous or obfuscated predictions when queried individually on any given input data. However, when their outputs are aggregated or combined, they yield accurate prediction vectors. This approach ensures that, while each model's standalone predictions are misleading and thus secure against theft from untrusted cloud adversaries, together they contribute to a correct and reliable ensemble prediction.

\subsection{Training Methodology} \label{stiq_train}
Training QNNs on cloud-based quantum hardware involves a multi-step process, combining classical and quantum computing techniques: 1) \textit{Design and Encode:} The first step involves designing the necessary parameterized quantum circuits for training. Classical data is encoded into these circuits using techniques such as angle encoding, amplitude encoding, etc. The designed circuit is then sent to the cloud provider. 2) \textit{Transpile and Map:} Upon receiving the circuit, it undergoes transpilation to match the basis gates supported by the cloud quantum hardware. This process also includes routing and mapping the logical qubits to the physical qubits of the quantum hardware, ensuring compatibility and optimization for execution. 3) \textit{Execute and Measure:} The circuit is executed on the quantum hardware. The qubits are then measured, and the expectation value, which serves as an output for the given input parameters, is calculated and returned. 4) \textit{Gradient Calculation:} To update the parameters of the quantum circuit effectively, the gradient of the parameters needs to be calculated. This involves several additional executions of the circuit or variations of it. The specific number of additional executions required depends on the technique used for gradient calculation, such as the parameter-shift rule or finite differences. 5) \textit{Parameter Optimization:} After calculating the gradients and determining the loss function, a classical optimizer is employed to update the circuit parameters. This optimization step is aimed at minimizing the loss function, thereby improving the model's performance on the training data. Steps 1-5 are iteratively executed until the QNN achieves the required accuracy or a predefined loss threshold is met.

In the conventional training of QNNs for image classification tasks, the process involves comparing the raw output vector $y$ from the QNN with the actual prediction vector $\hat{y}$. The loss $L$ is calculated using a classification loss function $C()$, represented as:
\begin{align*}
    L = C(y, \hat{y})
\end{align*}
This standard approach aims for each QNN to directly predict the correct output vector. However, our novel approach seeks to train multiple QNNs in parallel, where each QNN intentionally generates incorrect output vectors. The key innovation is that, when these outputs are combined, they yield the correct prediction vector. This method is particularly designed to safeguard against adversaries aiming to steal the model by making individual outputs misleading when considered in isolation.

\begin{figure*}[!t]
    \vspace{-2mm}
    \centering
    \includegraphics[width=0.95\textwidth]{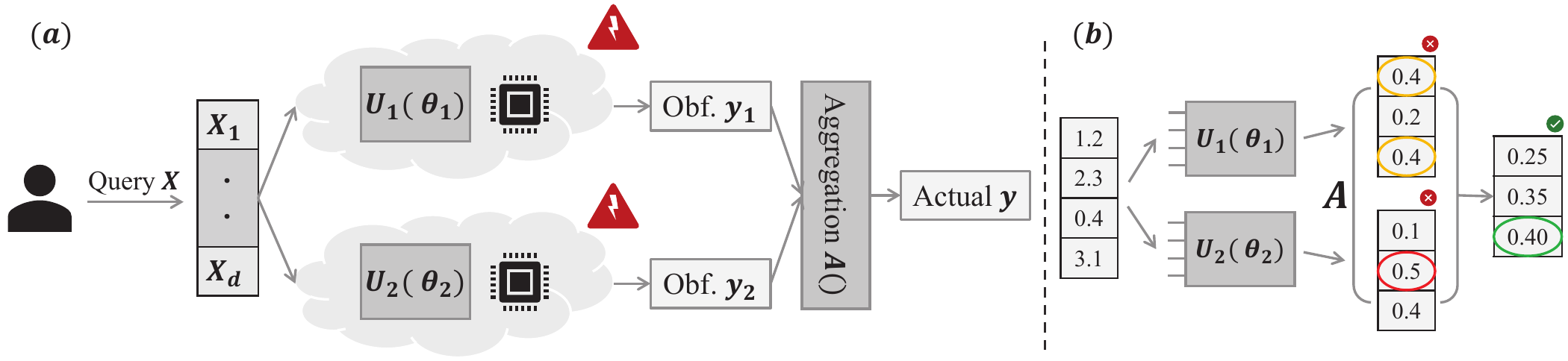}
    \vspace{-1mm}
    \caption{Figure depicting the STIQ inference methodology: (a) A user query \(X\) is processed by two cloud-hosted QNNs, resulting in obfuscated outputs \(y_1\) and \(y_2\). Locally, STIQ aggregates these outputs to produce the correct result \(y\) for the user. (b) Illustration of an example explaining how individual models may yield erroneous predictions, yet their combined output through STIQ accurately produces the correct output vector.} 
    \label{inference}
    \vspace{-4mm}
\end{figure*}

For this purpose, we introduce a unique training scheme for two QNNs, as shown in Fig. \ref{training}, which output vectors $y_1$ and $y_2$, respectively. These vectors are aggregated using a function $A()$, with the goal of matching the correct prediction vector $\hat{y}$. There can be several aggregation functions which can be used like sum, max, product followed by normalization. However, in this work we mainly focused on using $Mean()$ as our aggregation function $A()$. The classification loss $L_C$ for the aggregated output is thus:
\begin{align*}
    L_C = C(A(y_1, y_2), \hat{y})
\end{align*}
Now, to ensure that each QNN’s output is individually misleading and not directly useful for adversaries, we incorporate a divergence function $D()$. This function measures the divergence between $y_1$ and $y_2$, rewarding dissimilarity to encourage outputs that, while incorrect and diverse on their own, are complementary when aggregated. For instance, a function like Cosine Similarity, which yields a high value for identical vectors and a low value for dissimilar ones, can serve this purpose. Alternative loss functions, such as L1 or L2 loss, could also be adapted by adjusting their signs to ensure high values for similarity and low values for divergence.

Directly integrating the divergence loss \(L_D\) with the classification loss \(L_C\) might hinder the training effectiveness, as the model may struggle to simultaneously minimize both, potentially compromising the combined performance. To address this, we introduce a penalty factor \(\lambda\) within the range \(0, 1]\). This factor allows for the fine-tuning of the divergence loss’s impact on the overall training process. Therefore, the total loss \(L_T\) is formulated as:
\begin{align} \label{eq:1}
    &&L_T &= L_C + \lambda \cdot L_D&& \\
    \text{where,}&&L_C &= C(A(y_1, y_2), \hat{y})&& \notag \\
    &&L_D &= D(y_1, y_2)&& \notag\\
    &&\lambda &\in (0, 1]&& \notag
\end{align}
This formula balances the need for the QNNs to collectively produce accurate predictions while ensuring individual outputs are suitably divergent to deter model stealing, enabling controlled and effective training. During the training process, we optimize each QNN by utilizing the total loss, \(L_T\), in conjunction with a different optimizer tailored to the corresponding QNN.

\subsection{Inference Methodology}
Fig. \ref{inference}(a) shows the execution workflow of STIQ during the inference phase of QNNs hosted on the cloud. When a user wishes to utilize our models, they start by submitting a query image or data $X$. This query is concurrently forwarded to two QNN models, symbolized by the unitaries $U_1(\theta_1)$ and $U_2(\theta_2)$, which may be deployed across the same or different cloud providers, potentially untrusted. Each QNN model runs on the respective available hardware, generating output vectors $y_1$ and $y_2$, which are then returned to us. Crucially, these output vectors are obfuscated protecting the model from potential theft by any cloud-based adversaries. Subsequently, we employ a specific aggregation function, which was also used during the training phase, to merge the vectors $y_1$ and $y_2$. This aggregation yields the final corrected probability vector, which is then relayed back to the user. To demystify the process and clarify how the integration of two obfuscated vectors yields the accurate output vector, we provide a practical example.

Consider a 4-qubit QNN trained on a dataset with four input features and three output classes (labelled 0, 1, and 2) (Fig. \ref{inference}(b)). During inferencing, a user's input, represented as [1.2, 2.3, 0.4, 3.1], is processed by two separately trained QNNs $U_1(\theta_1)$ and $U_2(\theta_2)$. Assuming the correct class for this input is class-2, each QNN is trained to produce an obfuscated probability vector. The highest probability in each vector does not directly correspond to the correct class label, enhancing security against potential adversarial attacks. In the example shown, $U_1$ yields an output that ambiguously suggests either class-0 or class-2 could be correct, failing to definitively identify class-2. Conversely, $U_2$ incorrectly identifies class-1 as the most probable. By applying our aggregation function $A()$, which in this case computes the mean of the two vectors, we derive the corrected probability vector. This vector accurately identifies class-2 as the correct classification with high confidence.

\begin{figure*}[!t]
    \vspace{-7mm}
    \centering
    \includegraphics[width=\textwidth]{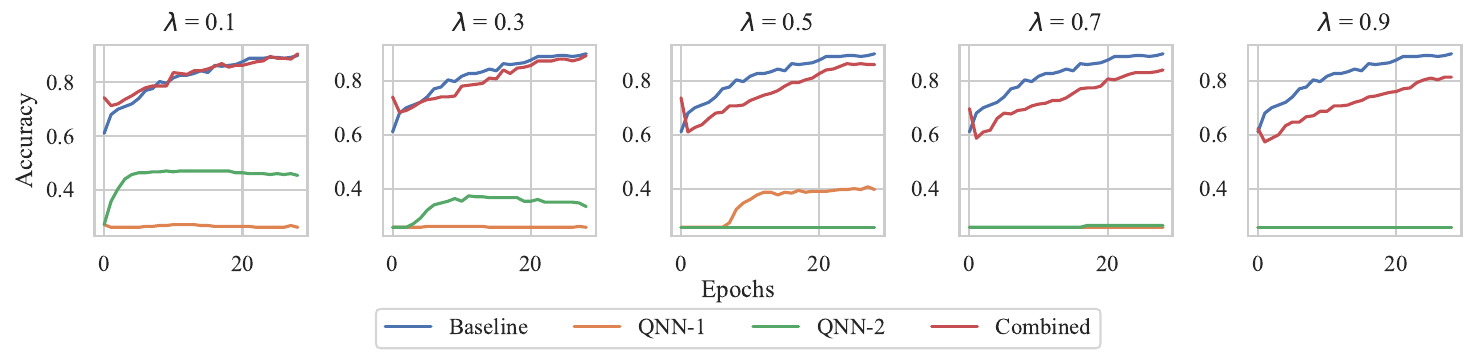}
    \vspace{-5mm}
    \caption{Plot demonstrating the test accuracies of different 4-qubit QNNs trained on the Fashion-4 dataset with varying penalties ($\lambda$). It is clear that while increasing the penalty enhances obfuscation, it also adversely affects the combined performance.} 
    \label{penalty_analysis}
    \vspace{-2mm}
\end{figure*}


\section{Evaluation}
\subsection{Setup} \label{setup}

\noindent \textbf{Device:} In our study, we conducted all noiseless experiments on Pennylane's \cite{bergholm2018pennylane} ``lightning.qubit" device using the "Adjoint" differentiation method to calculate the gradients. This method is known for its efficiency in simulating quantum circuits without noise. For our noisy inferencing, we used ``qiskit.aer" device. Given the long queue times and limited availability of real quantum devices and our goal to closely replicate the conditions of these noisy quantum devices, we opted to directly integrate noise models from actual quantum hardware. Specifically, we employed noise models from IBM\_Sherbrooke, IBM\_Cairo and IBM\_Hanoi and incorporated them to the "qiskit.aer" device to simulate the noise characteristics of these machines accurately. Considering that traditional methods for calculating gradients, such as backpropagation, are not viable on actual quantum devices, we utilized the Simultaneous Perturbation Stochastic Approximation (SPSA) method for gradient calculation in our QNN training. This approach produces noisy gradients, which are more indicative of the performance one would expect from real quantum hardware, thereby ensuring our experimental results are more aligned with practical quantum computing environments. We also conducted experiments on real quantum hardwares, specifically IBM\_{Sherbrooke, Brisbane and Kyoto}.

\begin{table}[!b]
    \vspace{-4mm}
    \centering
    \caption{Comparison of final test accuracy after training a 4-qubit QNN on Fashion-4 with varying penalties.}
    \label{tab_penalty}
    \vspace{0mm}
    \begin{tabular}{ccccc}
    \cmidrule(lr){1-5}
    Penalty    & Baseline & QNN-1 & QNN-2 & Combined \\
    \cmidrule(lr){1-5}
    $\lambda$ = 0.1     & \multirow{5}{*}{90.0}      & 25.9          & 45.3      & 90.6      \\
    $\lambda$ = 0.3     &       & 25.6           & 33.3      & 89.3      \\ 
    $\lambda$ = 0.5   &       & 39.6           & 25.6     & 86.0       \\ 
    $\lambda$ = 0.7   &       & 25.6           & 26.3      & 83.9        \\
    $\lambda$ = 0.9   &       & 25.6           & 25.6      & 81.3        \\
    \bottomrule
    \end{tabular}
    \vspace{-2mm}
\end{table}

\noindent \textbf{Training:} For evaluation, we used PQC-4 and 19 from \cite{sim2019expressibility} to build our QNNs (PQC-$x$ represents circuit-$x$ in \cite{sim2019expressibility}), initialized with random weights. In most of the experiments we used a 4-qubit or 8-qubit QNN, consistent with other QML works \cite{wang2022qoc, wang2022quantumnas}. The baseline QNN utilizes a generic training procedure that relies solely on the classification loss (i.e. CrossEntropyLoss()) to perform the required classification tasks. In contrast, QNN-1 and QNN-2 are trained concurrently using our STIQ methodology, which incorporates a combination of both classification and divergence losses, as explained in Section \ref{stiq_train}. For embedding classical features to their corresponding quantum state we use the angle encoding technique with $RZ$ gate and for measurement, we calculate the Pauli-Z basis expectation value over all the qubits. For training the QNNs we used CrossEntropyLoss() as our classification loss function and CosineSimilarity() as the divergence loss function (if not mentioned otherwise). The hyperparameters used for training are; Epochs: 30, learning\_rate ($\eta$): 0.01, batch\_size = 32 and optimizer: Adam. All training are done on an Intel Core-i7-12700H CPU with 40GB of RAM.

\noindent \textbf{Dataset:} Since NISQ devices struggle with large images due to limited qubits, high error rates and complex data encoding, similar to recent works \cite{wang2022qoc, fu2024quantumleak}, we conduct all experiments using a reduced feature set of MNIST, Fashion, Kuzushiji and Letters datasets with latent dimension $d = 8$ (from original 28$\times$28 image) generated using a convolutional autoencoder \cite{alam2021quantum}. Thus, for each dataset, we create a smaller 4-class dataset from these reduced feature sets i.e., MNIST-4 (class 0, 1, 2, 3), Fashion-4 (class 6, 7, 8, 9), Kuzushiji-4 (class 3, 5, 6, 9) and Letters-4 (class 1, 2, 3, 4) with each having 1000 samples (700 for training and 300 for testing). Since we use 4-qubit QNN models for training and each of these datasets is of dimension $d = 8$, we encode 2 features per qubit. Number of shots/trials is set to 1000 for all experiments.

\begin{table}[!b]
    \vspace{-2mm}
    \centering
    \caption{Final test accuracies after training a QNN on MNIST-4 with  different divergence loss functions. }
    \label{tab_loss}
    \vspace{0mm}
    \begin{tabular}{ccccc}
    \cmidrule(lr){1-5}
    Divergence Loss    & Baseline & QNN-1 & QNN-2 & Combined \\
    \cmidrule(lr){1-5}
    L1     & \multirow{3}{*}{91.6}      & 30.7          & 46.9      & 88.6      \\
    L2     &       & 29.4           & 22.6      & 81.6      \\ 
    CosineSimilarity &     & 31.9           & 40.3      & 91.3       \\ 
    \bottomrule
    \end{tabular}
    \vspace{-2mm}
\end{table}

\begin{figure}[!b]
        \vspace{-2mm}
        \centering 
        \includegraphics[width=\linewidth]{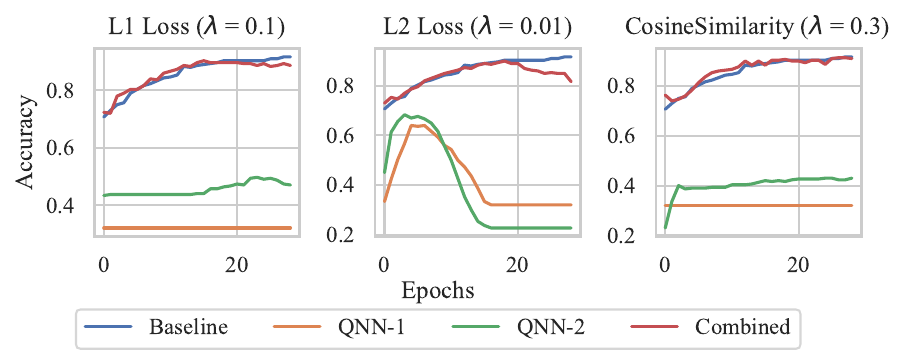}
        \vspace{-5mm}
        \caption{Test accuracies of 4-qubit QNNs trained on the MNIST-4 dataset with various divergence loss functions ($L_D$).}
        \label{loss_analysis}
        \vspace{-4mm}
\end{figure}

\begin{table*}[t]
    \vspace{-6mm}
    \centering
    \caption{Comparison of final test accuracy and loss after training a 8-qubit 2-layer QNN on different datasets for 30 epochs with learning rate $0.01$ and penalty ($\lambda$): 0.3. We can see that eventhough performance on individual QNNs are greatly obfuscated, performance of their combined outputs is similar to that of the baseline model.}
    \label{all_dataset}
    \vspace{-2mm}
    \begin{tabular}{ccccccccc}
    \cmidrule(lr){2-9}
    \multicolumn{1}{c}{} & \multicolumn{2}{c}{Baseline} & \multicolumn{2}{c}{QNN-1} & \multicolumn{2}{c}{QNN-2} & \multicolumn{2}{c}{Combined} \\
    \cmidrule(lr){1-9}
    Datasets    & Test Acc. (\%) & Test Loss & Test Acc. (\%) & Test Loss & Test Acc. (\%) & Test Loss & Test Acc. (\%) & Test Loss \\
    \cmidrule(lr){1-9}
    MNIST-4     & 98.3      & 0.099     & 22.6     & 5.346     & 37.6      & 2.915 & 98.3    &  0.137     \\
    Fashion-4     & 95.6      & 0.183      & 36.6      & 4.158     & 42.6      & 3.685 &  95.9   &  0.224     \\ 
    Kuzushiji-4   & 88.3      & 0.398      & 43.3     & 4.388     & 27.3      & 5.718 &  89.3    &  0.405     \\ 
    Letters-4   & 92.3      & 0.325      & 30.6    & 5.430     & 27.0      & 5.310 &  93.3      &  0.375     \\ 
    \bottomrule
    \end{tabular}
\end{table*}

\begin{figure*}[!t]
    \vspace{-4mm}
    \centering
    \includegraphics[width=0.95\textwidth]{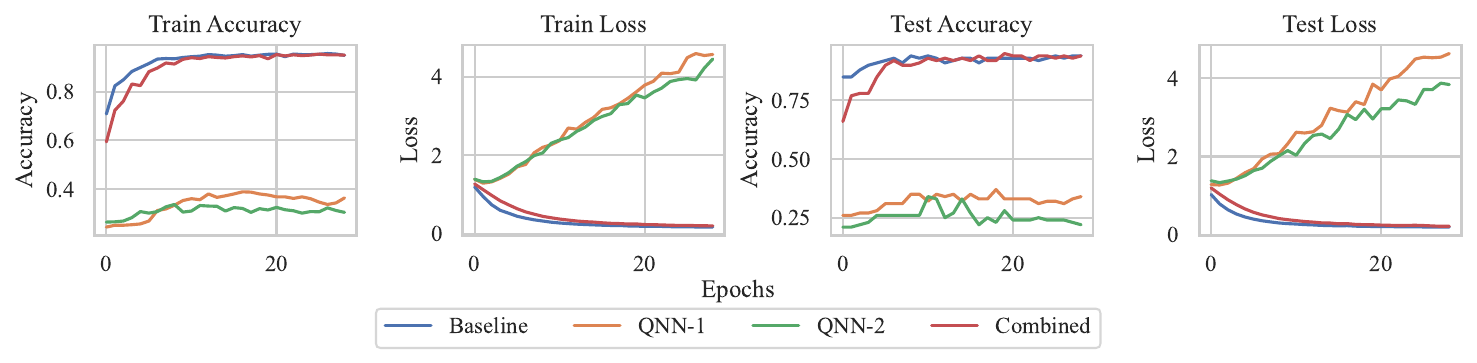}
    \vspace{-3mm}
    \caption{Performance of 8-qubit 2-layer QNN on Fashion-4 dataset. The plot clearly demonstrates that the combined performance of STIQ closely aligns with that of the baseline model, while the individual performances of QNN-1 and QNN-2 show decreased accuracy due to the intentional obfuscation introduced by our approach.} 
    \label{fashion_8q}
    \vspace{-4mm}
\end{figure*}

\begin{table}[!b]
    \vspace{-2mm}
    \centering
    \caption{Comparison of Obfuscation and Performance Gap (\(\Delta\)) between STIQ and Baseline QNN Across Different Datasets. Obfuscation is defined using the ratio \(\frac{\text{mean(QNN-1, QNN-2)}}{\text{Baseline}}\), indicating the average relative performance of the obfuscated QNNs compared to the Baseline. Performance \(\Delta\) is measured as \(\frac{\text{Combined}}{\text{Baseline}}\), representing the performance of the combined QNN outputs relative to the Baseline.}
    \label{tab:obf_datasets}
    \vspace{0mm}
    \begin{tabular}{ccccc}
    \cmidrule(lr){2-5}
     \multicolumn{1}{c}{} & \multicolumn{2}{c}{Obfuscation} & \multicolumn{2}{c}{Performance $\Delta$} \\
    \cmidrule(lr){1-5}
    Dataset    & Accuracy & Loss & Accuracy & Loss \\
    \cmidrule(lr){1-5}
    MNIST-4     & 0.3$\times$      & 41.3$\times$          & 1.0$\times$     & 1.37$\times$     \\
    Fashion-4     & 0.4$\times$      & 21.4$\times$           & 1.0$\times$      & 1.22$\times$     \\ 
    Kuzushiji-4   & 0.4$\times$     & 12.7$\times$           & 1.01$\times$      & 1.02$\times$       \\ 
    Letters-4   & 0.3$\times$      & 16.5$\times$           & 1.01$\times$      & 1.15$\times$        \\
    \bottomrule
    \end{tabular}
    \vspace{-2mm}
\end{table}

\subsection{Simulation Results}

\noindent \textbf{Penalty ($\lambda$) Analysis:} To achieve the goal of balancing obfuscation and performance in our QNN setup, we first need to determine the optimal value for the penalty parameter, $\lambda$, as introduced in Eq. \ref{eq:1}. This parameter is critical for adjusting the impact of the divergence loss $L_D$, in this case, CosineSimilarity, on the overall training process. Our objective is to find a $\lambda$ value that allows individual models to maintain a high degree of obfuscation without significantly compromising the combined performance of the models when compared to an unobfuscated baseline model. An analysis of this is illustrated in Fig. \ref{penalty_analysis} and Table \ref{tab_penalty}, which shows the test accuracy of a 4-qubit QNN model trained on the Fashion-4 dataset. The figure reveals that as $\lambda$ increases, enhancing the level of obfuscation by giving more weight to the divergence loss, the performance of the individual models, QNN-1 and QNN-2, begins to decline. This decrease in performance suggests that a high penalty on divergence loss can negatively affect the model's ability to optimize effectively, as the divergence loss begins to dominate over the classification loss. Based on the trends observed in the figure, it becomes clear that an optimal range for $\lambda$ exists. Specifically, for the CosineSimilarity divergence function, a $\lambda$ value within the range of $[0.1, 0.3]$ seems to be the optimal range. Within this range, the models achieve a desirable level of obfuscation (45-75\%) while still performing closely to the unobfuscated baseline model in terms of accuracy ($\leq 1\%$).

\noindent \textbf{Divergence Loss ($L_D$):} Next, we evaluate the effects of employing various divergence loss functions on the training and performance of 4-qubit QNNs on the MNIST-4 dataset. The divergence loss functions under consideration are L1 loss, L2 loss, and CosineSimilarity. The goal is to understand how each loss function influences the model's test accuracy, as depicted in Fig. \ref{loss_analysis} and Table \ref{tab_loss}. To ensure effective training of the models with each type of loss function, it was necessary to empirically determine suitable penalty values for each case. This step was crucial because the choice of divergence loss function and its corresponding penalty value directly affects the model's ability to learn. The analysis revealed a notable pattern: while initially, the models trained with L1 and L2 loss functions showed promising performance, their combined performance began to decline after reaching a certain number of epochs. This decline is is because of the divergence loss ($L_D$), which, beyond a certain point, starts to divert more from the overall learning process than the classification loss ($L_C$) contributes. Essentially, the models become too focused on minimizing $L_D$ at the expense of $L_C$, leading to poorer overall performance. In contrast, models utilizing CosineSimilarity as the divergence loss function exhibited a more stable training progression. This stability suggests that CosineSimilarity, unlike L1 or L2 loss, provides a balance that prevents the divergence loss from overwhelming the classification loss. Consequently, our findings indicate that CosineSimilarity is a preferable choice for the divergence loss function in the context of training QNNs, as it supports more consistent and effective learning.

\noindent \textbf{Different Datasets:} After determining the optimal divergence loss function and penalty value ($\lambda$) for achieving balanced performance, we train larger, more complex 8-qubit and 2-layer (PQC-19 \cite{sim2019expressibility}) QNNs. For this training, we use CosineSimilarity as our divergence loss function ($L_D$), with a penalty $\lambda$ set at 0.3, as identified through our earlier analysis. We conduct extensive training and testing on various datasets to evaluate the effectiveness of our proposed STIQ model under these conditions. The performance metrics, illustrated in Fig. \ref{fashion_8q}, provide a detailed comparison between the various models. The results depicted in the figure demonstrate that the combined performance of STIQ closely aligns with that of the baseline model, while the individual performances of QNN-1 and QNN-2 show decreased accuracy due to the intentional obfuscation introduced by our STIQ approach. Table \ref{all_dataset} shows the test performance of STIQ across various datasets. Although the test loss of the combined model is slightly higher across all datasets, the test accuracies indicate that the combined performance closely matches that of the unobfuscated baseline model. This observation is further supported by Table \ref{tab:obf_datasets}, which demonstrates the average obfuscation levels of QNN-1 and QNN-2 relative to the baseline QNN, as well as the performance gap (\(\Delta\)) between the combined and baseline models. On average, the accuracy and loss of the obfuscated QNNs are \(0.35\times\) and \(23\times\) those of the baseline QNN, respectively. However, the combined accuracy aligns closely (\(1\times\)) with that of the baseline, albeit with a \(1.19\times\) higher loss. These findings further showcases STIQ’s effectiveness in safeguarding QNNs from untrusted cloud providers.

\begin{table}[!t]
    \vspace{-2mm}
    \centering
    \caption{Test accuracies for 8-qubit QNNs trained on various datasets and infered across different hardware setups. The baseline, QNN-1, and QNN-2 models are tested on devices replicating the noise characteristics and basis gates of IBM\_Sherbrooke, IBM\_Hanoi, and IBM\_Cairo, respectively.}
    \label{tab:fake_infer_datasets}
    \vspace{0mm}
    \begin{tabular}{ccccc}
    \cmidrule(lr){1-5}
    Dataset    & Baseline & QNN-1 & QNN-2 & Combined \\
    \cmidrule(lr){1-5}
    MNIST-4     & 97.0      & 22.6          & 38.3      & 97.3     \\
    Fashion-4     & 95.0      & 37.3           & 40.1      & 95.0     \\ 
    Kuzushiji-4   & 86.6      & 44.0           & 27.3      & 87.6       \\ 
    Letters-4   & 91.0      & 37.0           & 21.0      & 91.0        \\
    \bottomrule
    \end{tabular}
    \vspace{-2mm}
\end{table}

\begin{figure}[!t]
    \vspace{0mm}
    \centering 
    \begin{subfigure}
        \centering 
        \includegraphics[width=0.8\linewidth]{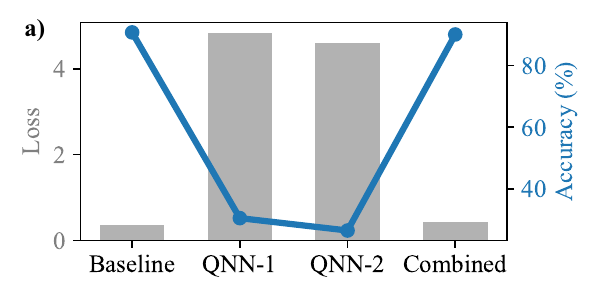}
        \vspace{2mm}
    \end{subfigure}
    \centering 
    \begin{subfigure}
        \centering
        \includegraphics[width=0.8\linewidth]{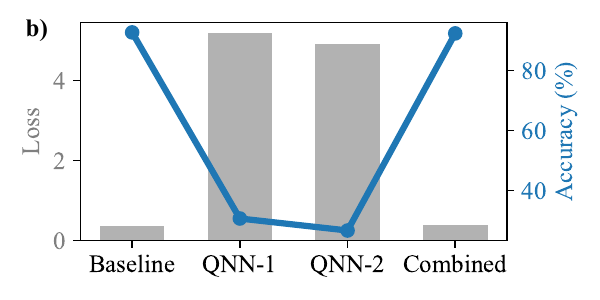}
    \end{subfigure}
    \vspace{-4mm}
    \caption{Test performance of 8-qubit QNNs trained on Letters-4 dataset in a noisy environment: (a) Performance across models on a single device with noise and basis gates similar to IBM\_Sherbrooke, and (b) Performance across different hardware setups, where the baseline, QNN-1, and QNN-2 are executed on devices mimicking the noise characteristics of IBM\_Sherbrooke, IBM\_Hanoi, and IBM\_Cairo respectively.}
    \label{fig:fake_inf}
    \vspace{-4mm}
\end{figure}

\noindent \textbf{Performance under Noise:} To evaluate the effectiveness of STIQ in noisy environments, we use our trained 8-qubit 2-layer QNN models on noiseless hardware and then test them on noisy hardware that simulates the noise models and basis gates of real quantum devices. We conducted tests using 300 images under two different setups: (i) all models were executed on a single device that mimics the noise model and basis gates of IBM\_Sherbrooke, and (ii) in a scenario that mirrors practical applications more closely, where each model—baseline, QNN-1, and QNN-2—was executed on different devices reflecting the noise characteristics and native gates of IBM\_Sherbrooke, IBM\_Hanoi, and IBM\_Cairo, respectively. This is a more realistic scenario since it helps mitigate the risk of reverse engineering by cloud providers, as it diversifies the hosting environments for the models. Table \ref{tab:fake_infer_datasets} presents the test results for models trained on various datasets and tested across different devices, showing that STIQ consistently matches the baseline performance with no degradation. Figure \ref{fig:fake_inf} illustrates the inference performance of all models, tested either on the a) same or on b) different devices. In both scenarios, the overall performance is comparable to the baseline, with individual models demonstrating significant obfuscation. This highlights STIQ's effectiveness in maintaining performance integrity under realistic noise conditions.

\subsection{Real Hardware Results}
Here, we evaluated the performance of STIQ on real quantum hardware by testing 4-qubit 2-layer QNNs (with PQC: circuit-1 \cite{sim2019expressibility}) on IBM\_Kyoto, as shown in Table \ref{tab:real_infer}. The results from the Letters-4 dataset indicate that the combined performance of the models was similar to the baseline, achieving similar accuracy and slightly higher loss. However, the performance of individual QNNs was significantly obfuscated, with an average accuracy reduction of 60\% and a loss nearly 6$\times$ that of the baseline and combined predictions. This demonstrates STIQ's effectiveness in a realistic scenario, where models are inferred on real quantum hardware via a cloud provider. 

\begin{table}[!t]
    \vspace{-2mm}
    \centering
    \caption{Performance evaluation of QNNs on \textbf{real quantum hardware}. The models were initially tested on IBM\_Kyoto. Subsequently they were also tested on different hardwares: Baseline on IBM\_Sherbrooke, QNN-1 on IBM\_Brisbane, and QNN-2 on IBM\_Kyoto.}
    \label{tab:real_infer}
    \vspace{0mm}
    \begin{tabular}{ccccc}
    \cmidrule(lr){1-5}
    \multicolumn{5}{c}{IBM\_Kyoto} \\
    \cmidrule(lr){1-5}
    Metric    & Baseline & QNN-1 & QNN-2 & Combined \\
    \cmidrule(lr){1-5}
    Accuracy     & 89.0      & 40.0          & 20.0     & 88.0     \\
    Loss     & 0.53      & 1.25           & 3.14      & 0.61     \\
    \cmidrule(lr){1-5}
    \multicolumn{5}{c}{IBM\_Sherbrooke $|$ IBM\_Brisbane $|$ IBM\_Kyoto} \\
    \cmidrule(lr){1-5}
    Metric    & Baseline & QNN-1 & QNN-2 & Combined \\
    \cmidrule(lr){1-5}
    Accuracy     & 90.0      & 40.0          & 20.0     & 90.0     \\
    Loss     & 0.34      & 2.77           & 3.28      & 0.42     \\ 
    \bottomrule
    \end{tabular}
    \vspace{-2mm}
\end{table}

We also considered another practical setting where each model was inferred on distinct quantum hardwares to enhance security by diversifying the hosting hardware for the models. Specifically, the baseline model was tested on IBM\_Sherbrooke, QNN-1 on IBM\_Brisbane, and QNN-2 on IBM\_Kyoto. The results, shown in Table \ref{tab:real_infer}, demonstrate that STIQ effectively maintains performance even across different systems. This indicates that STIQ not only mitigates adversarial threats but also preserves the integrity and confidentiality of QML models in cloud environments.

\subsection{Discussion}

\noindent \textbf{Scalability:} Proposed STIQ scales well with increase in the size of individual QNNs and dataset. In order to quantify this, we trained QNNs sized 4-qubit to 16-qubit each having 2-layers on a larger MNIST-10 dataset with 2000 samples with train:test of 70:30. We used SPSA for gradient calculation which is a zero-order optimization method known for it's efficiency and noise-robustness in training QML models when coupled with classical optimizers like Adam \cite{wiedmann2023empirical}. From Table \ref{tab:scalability}, we can clearly see that even with increase in size of QNN, STIQ is effectively able to highly obfuscated individual hosted QNNs. Interestingly, for QNNs trained on a larger dataset i.e. MNIST-10 in this case, STIQ actually outperforms the Baseline QNN by up to 6\%, suggesting that STIQ not only protects QNNs from cloud-based adversaries but also leads to superior performance in larger QNNs.

\noindent \textbf{Applicability to VQAs:} STIQ can be effectively adapted to enhance the security of other VQAs such as the Quantum Approximate Optimization Algorithm (QAOA) and the Variational Quantum Eigensolver (VQE) against cloud-based adversaries. The loss function (\(L\)), in problems like finding the MaxCut using QAOA or determining the ground state energy of a molecule using VQE, involves maximizing or minimizing the expectation value of a Hamiltonian (\(H\)). Consequently, (\(L\)) in these scenarios can be expressed as \(L = \langle H \rangle\). Using this formulation of \(L\), the STIQ training methodology can be applied to concurrently optimize two VQA circuits ($U_1, U_2$). Considering each circuit produces obfuscated expectation values, $\langle H_1 \rangle\ $ and $\langle H_2 \rangle\ $ independently and a divergence function $D()$, penalty $\lambda$ and aggregate function $A()$, the new loss function to optimize both $U_1$ and $U_2$ would be:
\[ L = A( \langle H_1 \rangle\ , \langle H_2 \rangle\\) + \lambda \cdot D(\langle H_1 \rangle\ , \langle H_2 \rangle\\)\]
Thus, STIQ is not merely confined to QNNs; it can be seamlessly extended to any VQA that relies on optimizing parameters based on a defined loss function.

\begin{table}[!t]
    \vspace{-2mm}
    \centering
    \caption{Test accuracies for different qubit 2-layer QNNs trained on larger MNIST-10 datasets using SPSA for gradient calculation. STIQ seems to outperform Baseline while significantly obfuscating individual QNNs.}
    \label{tab:scalability}
    \vspace{-2mm}
    \begin{tabular}{ccccc}
    \cmidrule(lr){1-5}
    Size    & Baseline & QNN-1 & QNN-2 & Combined \\
    \cmidrule(lr){1-5}
    4-Q      & 57.2      & 22.3          & 11.1      & 65.3     \\
    8-Q      & 69.1      & 8.33           & 20.5      & 75.5     \\ 
    12-Q     & 89.6      & 21.5           & 10.0      & 93.0       \\ 
    16-Q    & 85.5      & 13.8           & 14.3      & 90.5        \\
    \bottomrule
    \end{tabular}
    \vspace{-4mm}
\end{table}

\noindent \textbf{Ensemble Learning vs STIQ:} Although previous studies have explored the use of ensemble learning in quantum classifiers \cite{schuld2018quantum, silver2022quilt, qin2022improving}, the foundational approaches differ significantly from the proposed STIQ methodology. In earlier works, individual classifiers are trained to each make correct predictions independently, with their outputs then aggregated and weighted according to their respective accuracies. In contrast, STIQ employs a novel training procedure where individual classifiers are intentionally trained to make highly inaccurate predictions, yet their combined output results in correct predictions. Moreover, the primary motivation behind prior research was to develop quantum-based ensemble models that deliver robust performance without specifically addressing security concerns. Conversely, STIQ is designed fundamentally to protect individual classifiers from cloud-based adversarial attacks while still achieving optimal performance.

\begin{figure}[!t]
        \vspace{-2mm}
        \centering 
        \includegraphics[width=\linewidth]{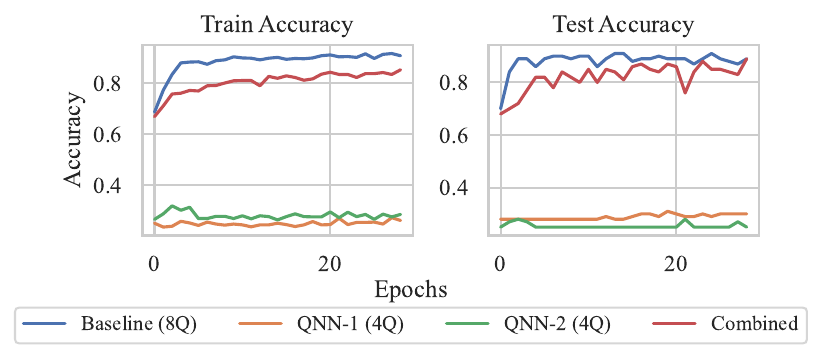}
        \vspace{-5mm}
        \caption{Train and test accuracies for different models on the Kuzushiji-4 dataset, comparing an 8-qubit baseline model with smaller 4-qubit models, QNN-1 and QNN-2.}
        \label{overhead_analysis}
        \vspace{-6mm}
\end{figure}

\noindent \textbf{Overhead Analysis:} Since STIQ involves training atleast 2 QNNs for each task, this increases the total computational overhead by approx 2$\times$ depending on the architecture of the individual QNNs. Using simpler models could help reduce this overhead but this might come at a cost of slightly decreased overall performance. We conducted experiments to evaluate this trade-off by comparing the performance of a baseline 8-qubit, 2-layer QNN (PQC: circuit-8 \cite{sim2019expressibility}) with that of a smaller 4-qubit, 2-layer QNNs (PQC: circuit-6 \cite{sim2019expressibility}). Figure \ref{overhead_analysis} presents the train and test accuracies for these models, demonstrating that while smaller models lead to an average performance degradation of about 2.5\%, they significantly reduce computational demands. This efficiency allows for training on smaller quantum hardwares, which can decrease the overall costs associated with training multiple models. Furthermore, while STIQ does increase the overall computational cost, it does not necessarily lead to a doubling of the training time. This is because the individual QNNs within the STIQ framework can be trained concurrently on different hardware platforms. As a result, if queue depths and gate execution times are similar, the total training time remains comparable to that of a baseline single-model training scenario.

\noindent \textbf{Security Analysis:} STIQ will provide security in following possible scenarios: (a) adversary gains access to trained QNN1 or QNN2: Although adversary will have full access to trained QNN1/QNN2, it will be useless since their individual performance is extremely poor. Adversary will need to train the model from scratch to make profit which defeats the purpose of stealth, (b) adversary gains access to both QNN1 and QNN2: Even in this scenario, an adversary cannot benefit because they won't know how to use the models effectively. If somehow the adversary learns that the outputs of individual models need to be combined, this knowledge is still insufficient. To effectively utilize the models, the adversary would also need to know the specific aggregate function used to combine the results, which is not revealed in the model.   

\section{Conclusion}
We address the risk of model theft when training or hosting QNNs on untrusted cloud platforms by introducing STIQ that utilizes ensemble techniques. STIQ involves hosting individual QNNs in such a manner that their outputs, when considered separately, are meaningless, effectively rendering them useless to potential attackers. However, when these outputs are aggregated locally, they yield the correct prediction, safeguarding the intellectual property embedded within the QNNs. We rigorously tested our technique across a variety of architectures, employing both noiseless simulators and those mimicking real device noise as well as directly on IBM's real quantum hardware. Our results demonstrate a significant degradation in the accuracy of individual models, while the accuracy of the combined output matches that of an unobfuscated QNN. 

\bibliographystyle{unsrt}
\bibliography{refs}

\end{document}